\documentclass[letter]{IEEEtran}
\IEEEoverridecommandlockouts
\usepackage{cite}
\usepackage{amsmath} 
\newcommand\ddfrac[2]{{\displaystyle\frac{\displaystyle #1}{\displaystyle #2}}}
\usepackage{amssymb,amsfonts}
\usepackage{algorithmic}
\usepackage{graphicx}
\usepackage{textcomp}
\usepackage{xcolor}
\def\BibTeX{{\rm B\kern-.05em{\sc i\kern-.025em b}\kern-.08em
    T\kern-.1667em\lower.7ex\hbox{E}\kern-.125emX}}
    
\definecolor{burgundy}{rgb}{0.5, 0.0, 0.13}  
\definecolor{brown(web)}{rgb}{0.65, 0.16, 0.16}
\definecolor{brightmaroon}{rgb}{0.76, 0.13, 0.28}
\definecolor{burntumber}{rgb}{0.54, 0.2, 0.14}
\definecolor{brickred}{rgb}{0.8, 0.25, 0.33}
\definecolor{cornellred}{rgb}{0.7, 0.11, 0.11}
\definecolor{darkcandyapplered}{rgb}{0.64, 0.0, 0.0}
\definecolor{darkred}{rgb}{0.55, 0.0, 0.0}
\definecolor{deepcarmine}{rgb}{0.66, 0.13, 0.24}

\usepackage{amsthm,amsmath,amssymb,mathtools,bm,etoolbox,float,hyperref}
\usepackage{graphicx,cite,cuted}
\usepackage{bigints}
\usepackage{caption}

\usepackage{stackengine}

\usepackage{ellipsis}
\usepackage{tkz-euclide}
\usepackage{tikz}
\usepackage{tikz-3dplot}
\usepackage[utf8]{inputenc}
\usepackage{pgfplots} 
\usepackage{pgfgantt}
\usepackage{pdflscape}
\pgfplotsset{compat=newest} 
\pgfplotsset{plot coordinates/math parser=false}
\pgfplotsset{every  tick/.style={black,},ylabel style={font=\tiny},xlabel style={font=\tiny},tick label style={font=\tiny},legend style= {font=\scriptsize},
minor x tick num=1,minor y tick num=1,xminorticks=true,yminorticks=true,}
  \newlength\fheight
\newlength\fwidth
 \usepackage{lipsum} 
 
 \usepackage{caption}
 \newtheorem*{remark}{Remark}
   
\begin{document}

\title{Rate and Power Adaptation for Multihop Regenerative Relaying Systems}

\author{\IEEEauthorblockN{Elyes Balti, \emph{Student Member, IEEE} and Brian K. Johnson, \emph{Senior Member, IEEE} }
\thanks{Elyes Balti is with the Wireless Networking and Communications Group, Department of Electrical and Computer Engineering, The University
of Texas at Austin, Austin, TX 78712 USA (e-mail: ebalti@utexas.edu).}%
\thanks{Brian K. Johnson is with the Department of Electrical and Computer Engineering, University
of Idaho, Moscow, ID 83844 USA e-mail: bjohnson@uidaho.edu .}
}

\maketitle

\begin{abstract}
In this work, we provide a global framework analysis of a multi-hop relaying systems wherein the transmitter (TX) communicates with the receiver (RX) through a set of intermediary relays deployed either in series or in parallel. Regenerative based relaying scheme is assumed such as the repetition-coded decoded-and-forward (DF) wherein the decoding is threshold-based. To reflect a wide range of fading, we introduce the generalized $H$-function (also termed as Fox-$H$ function) distribution model which enables the modeling of radio-frequency (RF) fading like Weibull and Gamma, as well as the free-space optic (FSO) such as the Double Generalized Gamma and M\'alaga fading. In this context, we introduce various power and rate adaptation policies based on the channel state information (CSI) availability at TX and RX. Finally, we address the effects of relaying topology, number of relays and fading model, etc, on the performance reliability of each link adaptation policy.
\end{abstract}


\section{Introduction}
Driven by the need for statistical models that better characterize fading, shadowing and atmospheric turbulences, the past few decades have seen a rise in the interest to develop more generalized statistical model that reflects wide range of distributions. This is often achieved by adding more parameters to already existing models which involves elementary as well as complicated mathematical functions. Mathematically speaking, generalized fading model is equivalent in deriving a general mathematical function that represents all possible elementary and complicated functions. Elliptic integrals, Zeta, Beta, Gamma, ERF, Mathieu $\&$ Spherical, and Bessels are known to be complicated and involved in many but limited number of probability distributions. To circumvent this limitation, Hypergeomteric family has been proposed as more general functions that can represent the previous mathematical functions, in particular, Meijer G-function \cite{wolfram}. Meijer G-function, although introduced back to 1936 by Cornelis Simon Meijer, has been recently resurrected to address the free space optical (FSO) fading models that are known to be complicated. However, the Fox H-function, introduced by Charles Fox in 1961, is a generalization of the Meijer G-function and therefore any statistical distributions can be expressed by Fox H-function \cite{bookfox}. In the literature, complicated fading distributions are derived in relaying networks and in particular for mixed radio frequency (RF)/FSO systems. In addition, relaying networks can cover different scenarios for regenerative such as repetition coded decode-and-forward \cite{cochannel,tractable} 
and non-regenerative like amplify-and-forward for fixed and variable relaying gains \cite{partial,icc}
In this work, we consider multihop regenerative relaying for serial and parallel topologies wherein all-active and selective relaying are considered as parallel deployment. We further consider the Fox $H$-function fading to represent the common RF and FSO fading models. Capitalizing on this, we evaluate the mutual information for different rate and power adaptation schemes. The remainder of the paper is organized as follows: Section II discusses the different topologies along with the derivations of the cumulative distribution function (CDF). Sections III and IV present the mutual information for different rate and power adaptation schemes relative to the availability of CSIR and CSIT. Section V presents the numerical simulation along with the analysis while the conclusive summaries and future direction are reported in Section VI.
\section{System Model}
In this section, we consider the analysis of serial and parallel relaying topologies wherein we derive the channel state information (CSI) statistics as a function of the number of relays $N$.
\subsection{Serial Relaying Topology}
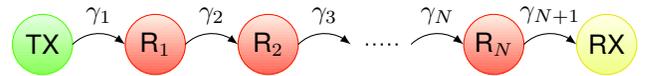
\begin{figure}[H]
\centering
\setlength\fheight{7.5cm}
\setlength\fwidth{7.5cm}
\usetikzlibrary{bending}
\begin{tikzpicture}
\draw [circle,top color=green!10, 
  bottom color=green!60!lime!70,
  draw=green!60!lime](-4,0) circle [radius=0.4] node {\sffamily{TX}};
  
  \draw [circle,top color=orange!10, 
  bottom color=red!80!orange!70,
  draw=red!80!orange](-2.5,0) circle [radius=0.4] node {\sffamily{R}$_1$};
  
    \draw [circle,top color=orange!10, 
  bottom color=red!80!orange!70,
  draw=red!80!orange](-1,0) circle [radius=0.4] node {\sffamily{R}$_2$};
  
      \draw [circle,top color=orange!10, 
  bottom color=red!80!orange!70,
  draw=red!80!orange](2,0) circle [radius=0.4] node {\sffamily{R}$_{N}$};
  
        \draw [circle,top color=yellow!10, 
  bottom color=yellow!50!lime!50,
  draw=yellow!50!lime](3.5,0) circle [radius=0.4] node {\sffamily{RX}};
  
    \draw [->,>=latex] (-3.6,0) to [bend left=45] node [above, sloped] (TextNode1) {$\gamma_1$} (-2.9,0);
    
    \draw [->,>=latex] (-2.1,0) to [bend left=45] node [above, sloped] (TextNode1) {$\gamma_2$} (-1.4,0);
    
    \draw [->,>=latex] (-0.6,0) to [bend left=45] node [above, sloped] (TextNode1) {$\gamma_3$} (0.1,0);
  
  \draw [dotted,thick] (.3,0) -- (0.75,0);
    \draw [->,>=latex] (0.9,0) to [bend left=45] node [above, sloped] (TextNode1) {$\gamma_N$} (1.6,0);
    
\draw [->,>=latex] (2.4,0) to [bend left=45] node [above, sloped] (TextNode1) {$\gamma_{N+1}$} (3.1,0);
  
\end{tikzpicture}
    \caption{Serial multihop relaying system consisting of a source $S$ communicating with a destination $D$ through the intermediary of $N$ relays. For $N$ serial relays, $S$ reaches out to $D$ through $N+1$ successive hops.}
     \label{serial-top}
\end{figure}
Fig.~\ref{serial-top} illustrates a set of relays deployed in series between the source and the destination. Unlike the parallel topology that will be discussed later, the signal can only reach the destination through one way defined by the serial relaying path. Besides, the serial relaying topology provides a sufficient power assistance in particular when the distance between $S$ and $D$ is very long. However, serial topology suffers from some limitations. Since the signal has only one way to reach $D$, the outage problem may arise when at least one relay becomes out of service or fails to decode the received signal. In this work, we assume regenerative relaying scheme wherein we consider the repetition coded decode-and-forward (DF) mode. In literature, conclusive summaries have been drawn about the performance of regenerative relaying which outperforms the non-regenerative scheme, however, regenerative relaying requires a threshold to decode the received signal. If the decoding threshold was not satisfied, the DF relay fails to forward the signal resulting in outage.

In the sequel analysis, we assume $N$ relays deployed in series between $S$ and $D$, i.e., there are $N+1$ hops between $S$ and $D$. We denote the $n$-th CSI by $\gamma_n$, $n=1,\ldots N+1$. The outage probability is defined as the probability when at least one CSI is less than a target threshold $\tau$. Mathematically, it is more straightforward to formulate the coverage probability and then express the cumulative distribution function (CDF) or the outage probability ($P_{\textsf{out}}({\scriptsize{\textsf{SNR}}},\tau) = 1 - P_{\textsf{coverage}}({\scriptsize{\textsf{SNR}}},\tau) $). Consequently, the CDF is expressed by
\begin{equation}
F_{\gamma}(\tau) = 1 - \mathbb{P}[\gamma_1\geq\tau,\gamma_2\geq\tau,\ldots,\gamma_{N+1}\geq\tau]    
\end{equation}
Assuming independence between the $N+1$ CSIs, the resulting CDF of the end-to-end CSI ($\gamma$) is expressed by
\begin{equation}
F_{\gamma}(\tau) = 1 - \prod_{n=1}^{N+1}\left(1-F_{\gamma_n}(\tau)\right)  
\end{equation}
where $F_{\gamma_n}(\cdot)$ is the CDF of the $n$-th CSI.
\subsection{Parallel Relaying Topology}
\begin{figure}[H]
\centering
\setlength\fheight{7.5cm}
\setlength\fwidth{7.5cm}
\usetikzlibrary{bending}
\begin{tikzpicture}
\draw [circle,top color=green!10, 
  bottom color=green!60!lime!70,
  draw=green!60!lime](-4,0) circle [radius=0.4] node {\sffamily{TX}};

    \draw [circle,top color=orange!10, 
  bottom color=red!80!orange!70,
  draw=red!80!orange](-0.25,0) circle [radius=0.4] node {\sffamily{R}$_n$};
  
        \draw [circle,top color=orange!10, 
  bottom color=red!80!orange!70,
  draw=red!80!orange](-.25,1.5) circle [radius=0.4] node {\sffamily{R}$_{1}$};
  
      \draw [circle,top color=orange!10, 
  bottom color=red!80!orange!70,
  draw=red!80!orange](-.25,-1.5) circle [radius=0.4] node {\sffamily{R}$_{N}$};
  
        \draw [circle,top color=yellow!10, 
  bottom color=yellow!50!lime!50,
  draw=yellow!50!lime](3.5,0) circle [radius=0.4] node {\sffamily{RX}};
  
    \draw [->,>=latex] (-3.6,0) to [] node [above, sloped] (TextNode1) {$\gamma_{1(1)}$} (-.65,1.5);
    
    \draw [->,>=latex] (-3.6,0) to [] node [above, sloped] (TextNode1) {$\gamma_{1(n)}$} (-.65,0);
    
    \draw [->,>=latex] (-3.6,0) to [] node [above, sloped] (TextNode1) {$\gamma_{1(N)}$} (-.65,-1.5);
    
    \draw [->,>=latex] (.17,1.5) to [] node [above, sloped] (TextNode1) {$\gamma_{2(1)}$} (3.1,0);
  
      \draw [->,>=latex] (.17,0) to [] node [above, sloped] (TextNode1) {$\gamma_{2(n)}$} (3.1,0);
      
    \draw [->,>=latex] (.17,-1.5) to [] node [above, sloped] (TextNode1) {$\gamma_{2(N)}$} (3.1,0);

  \draw [dotted,thick] (-.25,1) -- (-.25,.5);
  \draw [dotted,thick] (-.25,-.5) -- (-.25,-1);
  
\end{tikzpicture}
    \caption{Dualhop relaying system consisting of a source $S$ communicating with a destination $D$ through the intermediary of $N$ parallel relays.}
     \label{parallel-top}
\end{figure}
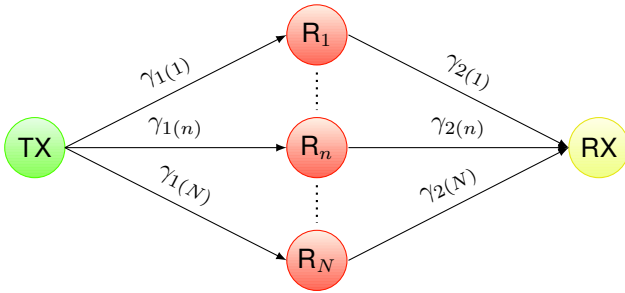
Fig.~\ref{parallel-top} illustrates a wireless communication system wherein $S$ and $D$ communicates through the intermediary of $N$ parallel relays. Unlike serial relaying topology, the signal can reach the destination through different ways which may decrease the outage occurence in particular when some relays fail to decode the received signal. In the literature, parallel relaying topology has been widely investigated, in particular, relaying protocols to define the signal routing between $S$ and $D$. In this work, we consider two relaying protocols as follows.
\subsubsection{All-Active Relaying} In this scenario, the signal reaches $S$ through the $N$ relays, i.e., the $N$ relays are simultaneously active to forward to the destination. Although this protocol results in the highest CSI at $D$, it requires not only huge power consumption but also it may arise the problem of synchronization at the receiver. 

Given that the relays are employing the DF scheme, the CSI of the $n$-th branch ($n=1,\ldots,N$) is given
\begin{equation}
    \gamma_n = \min(\gamma_{1(n)},\gamma_{2(n)}).
\end{equation}
Consequently, the output CSI at $D$ is the sum of all the CSIs of the $N$ links. The CDF of the end-to-end CSI is given by
\begin{equation}
F_{\gamma}(\tau) = \mathbb{P}\left[\sum_{n=1}^{N}\gamma_n\leq \tau\right].   
\end{equation}
\subsubsection{Selective Relaying}
In this scenario, the signal reaches the destination through one relay that is selected following a given protocol. In this context, we propose to select the relay/branch with the highest end-to-end CSI as follows
\begin{equation}
\gamma = \max\left( \min\left(\gamma_{1(1)},\gamma_{2(1)}\right),\ldots,\min\left(\gamma_{1(N)},\gamma_{2(N)}\right)  \right).    
\end{equation}
After some mathematical manipulations, the CDF of the end-to-end SNR is given by
\begin{equation}
\begin{split}
F_{\gamma}(\tau) =& \left(1 - \prod_{n=1}^N (1-F_{\gamma_{1(n)}}(\tau)) \right)\\&\times~\left(1 - \prod_{m=1}^N (1-F_{\gamma_{2(m)}}(\tau)) \right).     
\end{split}
\end{equation}

\subsection{H-function fading model}
This fading is considered as a generalized model of the common fading channels since it involves the famous Fox's $H$-function. It said that a given random variable $\gamma$ follows the unified Fox's $H$-function distribution if its PDF is expressed as follows \cite[Sec.~(4.1)]{fox} 
\begin{align}
    f_{\gamma}(\gamma) = \kappa H_{p,q}^{m,n} \Bigg(\delta\gamma ~\bigg|~\begin{matrix} (a_j,~A_j)_{j=1:p} \\ (b_j,~B_j)_{j=1:q}  \end{matrix} \Bigg), \gamma > 0 
\end{align}
where the parameters $\delta > 0$, and the constant $\kappa$ are chosen to satisfy $\int\limits_0^{+\infty} f_{\gamma}(\gamma)d\gamma = 1$. The univariate $H$-function, $H_{p,q}^{m,n}(\cdot)$, is defined by \cite{fox1}
\begin{align}
&H_{p,q}^{m,n} \Bigg(x ~\bigg|~\begin{matrix} (a_j,~A_j)_{j=1:p} \\ (b_j,~B_j)_{j=1:q}  \end{matrix} \Bigg) = \frac{1}{2\pi i}\\&\times~\int\limits_{\mathcal{C}}\ddfrac{\prod_{j=1}^{m}\Gamma(b_j+B_js) \prod_{j=1}^{n}\Gamma(1-a_j-A_js)}{\prod_{j=n+1}^p \Gamma(a_j+A_js)\prod_{j=m+1}^{q}\Gamma(1-b_j-B_js)}x^{-s}ds,  
\end{align}
Since Fox's $H$-function is generalized, common fading models for RF and FSO channels can be expressed in terms of this function. Special cases of the $H$-function distribution are summarized in Table \ref{tab1}.
\begin{remark}
Due to the limited number of pages for the letter, we provide the closed-form expressions of the coverage probability and the mutual information in the journal extension of this letter.
\end{remark}
\begin{table*}[t]
\renewcommand{\arraystretch}{1.3}
\centering
\captionsetup{font=scriptsize}
\caption{Common Fading Models}
\label{tab1}
\centering
\begin{tabular}{lc}
    \hline
    \hline
    \textbf{Fading Model}  &  \textbf{Probability Density Function}\\
    \hline\\

    \textbf{Exponential} \cite[Eq.~(2.7)]{simon} & $ \ddfrac{1}{\overline{\gamma}} H_{0,1}^{1,0} \Bigg(\ddfrac{\gamma}{\overline{\gamma}} ~\bigg|~\begin{matrix} - \\ (0,~1)  \end{matrix} \Bigg)$,\\

   \textbf{Gamma} \cite[Eq.~(2.21)]{simon}  & $\ddfrac{m}{\Gamma(m)\overline{\gamma}} H_{0,1}^{1,0} \Bigg(\ddfrac{m}{\overline{\gamma}}\gamma ~\bigg|~\begin{matrix} - \\ (m-1,~1)  \end{matrix} \Bigg) $,\\
   
    
    
    \textbf{Weibull} \cite[Eq.~(2.27)]{simon} & $\ddfrac{\omega}{\overline{\gamma}}H_{0,1}^{1,0} \Bigg(\ddfrac{\omega}{\overline{\gamma}}\gamma ~\bigg|~\begin{matrix} - \\ (1-1/\kappa,~1/\kappa)  \end{matrix} \Bigg)$,\\
    
     \textbf{Generalized Gamma} \cite{gam} & $\ddfrac{\beta}{\Gamma(m)\overline{\gamma}} H_{0,1}^{1,0} \Bigg(\ddfrac{\beta}{\overline{\gamma}}\gamma ~\bigg|~\begin{matrix} - \\ (m-1/\xi,~1/\xi)  \end{matrix} \Bigg) $,\\
     
         \textbf{Weibull-Gamma} \cite[Eq.~(4)]{wg} & $ \ddfrac{\beta \gamma^{\beta-1}}{\Gamma(\alpha)} \left(\frac{\Gamma(1+\frac{2}{\beta})}{\Omega}\right)^{\frac{\beta}{2}}\frac{\kappa^{1/2}\lambda^{\alpha - \frac{\beta + 1}{2}   }}{(2\pi)^{\ddfrac{\lambda+\kappa}{2}-1}}
G_{0,\kappa+\lambda}^{\kappa+\lambda,0} \Bigg(\frac{(\gamma^2\Gamma(1+2/\beta))^{\lambda}}{\Omega^\lambda \kappa^\kappa \lambda^\lambda} ~\bigg|~\begin{matrix} - \\ b_{\kappa+\lambda}\end{matrix} \Bigg)  $,\\
     
    \textbf{Gamma-Gamma} \cite[Eq.~(2)]{mod} & $\ddfrac{\xi^2}{\Gamma(\alpha)\Gamma(\beta)\gamma} H_{1,3}^{3,0} \Bigg(\ddfrac{(h\alpha\beta)^r}{\mu_r}\gamma ~\bigg|~\begin{matrix} (\xi^2+1,~1/r) \\ (\xi^2,~1/r),~(\alpha,~1/r),~(\beta,~1/r)\end{matrix} \Bigg)$, \\
    \textbf{Double Generalized Gamma} \cite[Eq.~(32)]{nasab} & $\ddfrac{ p^{m_2+1/2}q^{m_1-1/2}(2\pi)^{1-\frac{p+q}{2}}}{\alpha_2^{-1}\Gamma(m_1)\Gamma(m_2)\gamma} H_{p+q+\alpha_2p,\alpha_2p}^{0,p+q+\alpha_2p} \Bigg( \left( \left(\ddfrac{q\Omega_1}{m_1}\right)^{\frac{q}{p}}  \left( \ddfrac{p\Omega_2}{m_2}\right)\right)^{\frac{r}{\alpha_2}}  (A_0I_l)^r \ddfrac{\mu_r}{\gamma} ~\bigg|~\begin{matrix} (\kappa_1,~\frac{1}{r}) \\ (\kappa_2,~\frac{1}{r})\end{matrix} \Bigg) $,    \\
    \textbf{M\'alaga} \cite[Eq.~(51)]{egypt} & 
$\begin{cases} \ddfrac{A}{2}  \sum\limits_{n=1}^{\infty} a_nH_{0,2}^{2,0} \Bigg(\ddfrac{\alpha}{I}\gamma ~\bigg|~\begin{matrix} - \\ \left(\ddfrac{\nu+\alpha+n}{2},~1\right),~\left(\ddfrac{-\nu+\alpha+n}{2},~1\right)  \end{matrix} \Bigg),     &\mbox{if } \beta~ \mbox{is non-integer}  \\ 
\ddfrac{A}{2}  \sum\limits_{n=1}^{\beta} a_n H_{0,2}^{2,0}\Bigg(\ddfrac{\alpha\beta}{I\beta+\Omega^{'}}\gamma ~\bigg|~\begin{matrix} - \\ \left(\ddfrac{\nu+\alpha+n}{2},~1\right),~\left(\ddfrac{-\nu+\alpha+n}{2},~1\right)  \end{matrix} \Bigg),     & \mbox{if } \beta ~\mbox{is an integer} \end{cases} $
    \\\\
\hline\hline
\end{tabular}
\end{table*}
\section{Channel Capacity under CSIR}
In this section, we consider the systems with knowledge of CSI at the receiver, i.e., the ergodic capacity and effective capacity.  
\subsection{Optimal Rate Adaptation}
This rate adaptation transmission policy assumes that the transmit power is kept constant which is also called the ergodic capacity which is defined as
\begin{equation}
 \mathcal{I}({\scriptsize{\textsf{SNR}}}) = \frac{1}{2}\mathbb{E}[\log_2(1+\gamma)]   
\end{equation}
This expression can be transformed by integration by parts into
\begin{equation}
\mathcal{I}({\scriptsize{\textsf{SNR}}}) = \frac{1}{2\log(2)}\int\limits_0^{+\infty}  \ddfrac{1-F_\gamma(\gamma)}{1+ \gamma} d\gamma 
\end{equation}
\subsection{Effective Channel Capacity}
Modern radio systems such UMTS and LTE aim at supporting a set of services such as messaging, voice calls, videos sharing, etc which require a certain predefined quality-of-service (QoS) to be satisfied. Particularly, QoS metrics are commonly the delay, data rate and signal-to-interference-plus-noise ratio (SINR), etc. Authors in [53] introduced the concept of effective capacity as 
\begin{equation}
\mathcal{I}({\scriptsize{\textsf{SNR}}})  = -\frac{1}{\delta}  \log\left(\int\limits_0^{+\infty}\ddfrac{f_\gamma(\gamma)}{(1+\gamma)^{\frac{\delta}{2\log(2)}}}d\gamma  \right) 
\end{equation}
where $\delta = \phi B T_f$, $\phi, B$ and $T$ being the QoS exponent, bandwidth and fading block/frame length, respectively. Smaller values of $\phi$ correspond to slow decaying rate and looser QoS constraints while larger values correspond to fast decaying rate with more stringent QoS constraints. 
\section{Channel Capacity under CSIT and CSIR}
In this section, we consider a relaying systems wherein we assume the full knowledge of CSI at the transmitter and the receiver.

\subsection{Channel Inversion with Fixed Rate}
This technique offers an easy implementation to maintain a constant SNR at the receiver with fixed rate modulation and fixed code design or channel inversion with fixed rate (CIFR). Under this policy, the rate is expressed as
\begin{equation}
\mathcal{I}({\scriptsize{\textsf{SNR}}})  = \frac{1}{2\log(2)}\log\left(1+\left(\int\limits_0^{+\infty} \ddfrac{1}{\gamma}f_\gamma(\gamma)d\gamma\right)^{-1}  \right)  
\end{equation}
Besides, the CIFR technique may struggle large rate penalties in severe fading conditions as compared to other techniques. To address this shortcoming, a modified inversion which inverts the channel fading above a predetermined truncated fade $\gamma_0$ (TCIFR scheme) is often used.
\subsection{Truncated Channel Inversion with Fixed Rate}
The channel capacity based on the truncated channel inversion with fixed rate (TCIFR) policy is defined as
\begin{equation}
\mathcal{I}({\scriptsize{\textsf{SNR}}})  =\frac{1}{2\log(2)} \log\left(1+\left(\int\limits_{\gamma_0}^{+\infty} \ddfrac{1}{\gamma}f_\gamma(\gamma)d\gamma\right)^{-1}  \right) \mathbb{P}[\gamma \geq \gamma_0]   
\end{equation}
\subsection{Optimal Power and Rate Adaptation}
The optimal power and rate adaptation (OPRA) method achieves the highest possible rate with CSI by employing a multiplexed multiple codebook design to match the transmission power and rate of the system. The channel capacity for a system employing the OPRA technique is defined by
\begin{equation}\label{OPRA}
\mathcal{I}({\scriptsize{\textsf{SNR}}}) = \frac{1}{2}\int\limits_{\gamma_0}^{+\infty}\log_2\left(\frac{\gamma}{{\gamma_0}} \right)f_\gamma(\gamma)d\gamma    
\end{equation}
where $\gamma_0$ is the optimal cut-off SNR below which data transmission is not allowed. Consequently, $\gamma_0$ must satisfy the following requirement
\begin{equation}\label{condition}
\int\limits_{\gamma_0}^{+\infty} \left(\frac{1}{\gamma_0} - \frac{1}{\gamma}\right)f_\gamma(\gamma)  = 1    
\end{equation}
Using integration by parts, (\ref{OPRA}) becomes
\begin{equation}
\mathcal{I}({\scriptsize{\textsf{SNR}}})  = \frac{1}{2\log(2)}\int\limits_{\gamma_0}^{+\infty} \frac{1-F_\gamma(\gamma)}{\gamma}d\gamma
\end{equation}
\section{Numerical Results}
In this section, we investigate and verify the validity of the analytical expression derived in the preceding Sections.
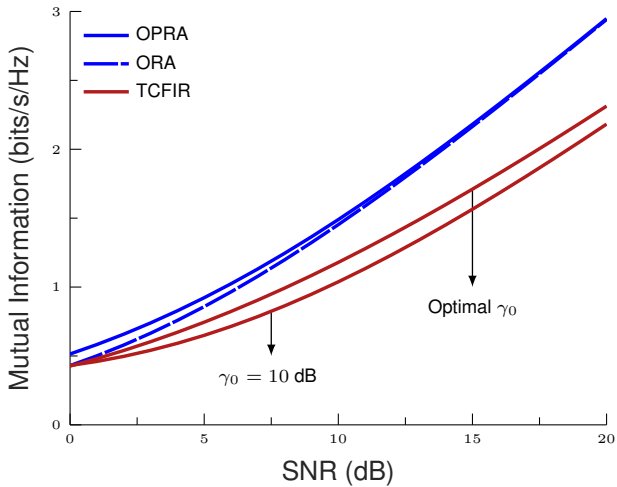
\begin{figure}[H]
\centering
\setlength\fheight{5.5cm}
\setlength\fwidth{7.5cm}
%
%
\definecolor{mycolor1}{rgb}{0.00000,0.44700,0.74100}%
\definecolor{mycolor2}{rgb}{0.85000,0.32500,0.09800}%
\definecolor{mycolor3}{rgb}{0.92900,0.69400,0.12500}%
\definecolor{mycolor4}{rgb}{0.49400,0.18400,0.55600}%
\begin{tikzpicture}

\begin{axis}[%
width=0.951\fwidth,
height=\fheight,
at={(0\fwidth,0\fheight)},
scale only axis,
xmin=0,
xmax=20,
xlabel style={font=\color{white!15!black}},
xlabel={\textsf{SNR (dB)}},
ymin=0,
ymax=3,
ylabel style={font=\color{white!15!black}},
ylabel={\textsf{Mutual Information (bits/s/Hz)}},
axis background/.style={fill=white},
axis x line*=bottom,
axis y line*=left,
legend style={at={(0.25,0.99)},legend cell align=left, align=left, draw=white!15!black,draw=none,fill=none}
]

\draw [-latex,black,line width=.5pt] (7.5,.82) to (7.5,.5);

\draw [-latex,black,line width=.5pt] (15,1.7) to (15,1);

\node[right, align=left, rotate=0]
at (axis cs:5,.35) {\scriptsize{\textsf{ $\gamma_0=10$ dB}}};

\node[right, align=left, rotate=0]
at (axis cs:13,.85){\scriptsize{\textsf{Optimal $\gamma_0$}}};

\addplot [color=blue, line width=1.3pt]
  table[row sep=crcr]{%
0	0.514269462679739\\
1	0.5828426036252\\
2	0.657870614784668\\
3	0.739477958792908\\
4	0.827715401821509\\
5	0.922556440366274\\
6	1.02389695199491\\
7	1.13155822107687\\
8	1.24529319834855\\
9	1.3647955652059\\
10	1.48971093266661\\
11	1.61964934651887\\
12	1.75419821475137\\
13	1.89293482134215\\
14	2.03543772555685\\
15	2.18129653681113\\
16	2.33011976810317\\
17	2.4815406738685\\
18	2.63522114609563\\
19	2.79085386911245\\
20	2.94816300129847\\
};
\addlegendentry{\textsf{OPRA}}

\addplot [color=blue,dash pattern={on 13pt off 1pt on 0pt off 0pt}, line width=1.3pt]
  table[row sep=crcr]{%
0	0.430173691135443\\
1	0.500925904125906\\
2	0.57914313651813\\
3	0.664818351665208\\
4	0.757837135225122\\
5	0.857987092533702\\
6	0.964971175354941\\
7	1.07842356284194\\
8	1.19792671467302\\
9	1.32302835725812\\
10	1.45325740421298\\
11	1.58813809677556\\
12	1.72720193405031\\
13	1.8699972146889\\
14	2.01609621411504\\
15	2.16510016720017\\
16	2.31664231902001\\
17	2.47038935327787\\
18	2.62604151787046\\
19	2.78333175631984\\
20	2.94202411684173\\
};
\addlegendentry{\textsf{ORA}}

\addplot [color=cornellred,line width=1.3pt]
  table[row sep=crcr]{%
0	0.426502386970889\\
1	0.480685445750347\\
2	0.539586925744779\\
3	0.603268366043102\\
4	0.671746775843574\\
5	0.74499522083186\\
6	0.822945157638622\\
7	0.905490294598005\\
8	0.99249163078427\\
9	1.08378324910777\\
10	1.17917842214898\\
11	1.27847562791475\\
12	1.38146415263802\\
13	1.48792905871348\\
14	1.59765539878615\\
15	1.71043164215823\\
16	1.82605235097095\\
17	1.94432016907157\\
18	2.06504723002923\\
19	2.18805606661672\\
20	2.31318012072368\\
};
\addlegendentry{\textsf{TCFIR}}

\addplot [color=cornellred, line width=1.3pt]
  table[row sep=crcr]{%
0	0.430749782456618\\
1	0.460242196580818\\
2	0.496973452114353\\
3	0.540814047771285\\
4	0.591687931100994\\
5	0.649526045693339\\
6	0.714233410320139\\
7	0.785667960779067\\
8	0.86362969792747\\
9	0.94785864348025\\
10	1.03803986914951\\
11	1.13381362381022\\
12	1.23478848555893\\
13	1.3405555818576\\
14	1.45070224264312\\
15	1.56482390412947\\
16	1.68253358653094\\
17	1.80346870017568\\
18	1.92729531211181\\
19	2.05371020032615\\
20	2.18244115338423\\
};

\end{axis}
\end{tikzpicture}%
    \caption{Channel capacity for different adaptive transmission schemes. Note that serial relaying topology has been assumed with 2 relays. Channel capacity for different adaptive transmission schemes. Note that selective relaying mode has been adopted with 3 relays. We assume the Gamma-Gamma FSO fading with strong turbulences ($\alpha=2.902,~\beta=2.51$) and severe pointing errors ($\xi = 1.1$) under heterodyne detection.}
    \label{figure1}
\end{figure}
In Fig.~(\ref{figure1}), we investigate the capacity of the system under different adaptive transmission schemes. It can be deduced that the mutual information under OPRA scheme performs better than the rest of the schemes, especially at low SNR, and this is in line with the definition of OPRA capacity as the highest achievable mutual information.

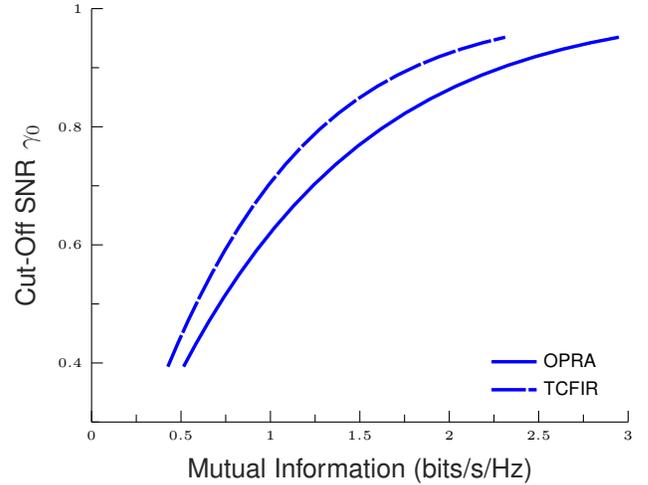
\begin{figure}[H]
\centering
\setlength\fheight{5.5cm}
\setlength\fwidth{7.5cm}
%
%
\definecolor{mycolor1}{rgb}{0.00000,0.44700,0.74100}%
\definecolor{mycolor2}{rgb}{0.85000,0.32500,0.09800}%
\begin{tikzpicture}

\begin{axis}[%
width=0.951\fwidth,
height=\fheight,
at={(0\fwidth,0\fheight)},
scale only axis,
xmin=0,
xmax=3,
xlabel style={font=\color{white!15!black}},
xlabel={\textsf{Mutual Information (bits/s/Hz)}},
ymin=0.3,
ymax=1,
ylabel style={font=\color{white!15!black}},
ylabel={\textsf{Cut-Off SNR}$~\gamma_0$},
axis background/.style={fill=white},
axis x line*=bottom,
axis y line*=left,
legend style={at={(0.97,0.03)}, anchor=south east, legend cell align=left, align=left, draw=white!15!black,draw=none}
]
\addplot [color=blue, line width=1.3pt]
  table[row sep=crcr]{%
0.514269462679739	0.393773845045118\\
0.5828426036252	0.432122349320114\\
0.657870614784668	0.471430465572848\\
0.739477958792908	0.511271175958566\\
0.827715401821509	0.551186950536065\\
0.922556440366274	0.590708160777954\\
1.02389695199491	0.629373226677489\\
1.13155822107687	0.66674881110626\\
1.24529319834855	0.702448271327248\\
1.3647955652059	0.73614673429037\\
1.48971093266661	0.767591564249833\\
1.61964934651887	0.796607567699445\\
1.75419821475137	0.82309692258988\\
1.89293482134215	0.847034407881696\\
2.03543772555685	0.868458948737302\\
2.18129653681113	0.887462728355275\\
2.33011976810317	0.904179145694992\\
2.4815406738685	0.918770756582008\\
2.63522114609563	0.931418084985431\\
2.79085386911245	0.942309896861698\\
2.94816300129847	0.951635245331063\\
};
\addlegendentry{\textsf{OPRA}}

\addplot [color=blue,dash pattern={on 13pt off 1pt on 0pt off 0pt}, line width=1.3pt]
  table[row sep=crcr]{%
0.426502386970889	0.393773845045118\\
0.480685445750347	0.432122349320114\\
0.539586925744779	0.471430465572848\\
0.603268366043102	0.511271175958566\\
0.671746775843574	0.551186950536065\\
0.74499522083186	0.590708160777954\\
0.822945157638622	0.629373226677489\\
0.905490294598005	0.66674881110626\\
0.99249163078427	0.702448271327248\\
1.08378324910777	0.73614673429037\\
1.17917842214898	0.767591564249833\\
1.27847562791475	0.796607567699445\\
1.38146415263802	0.82309692258988\\
1.48792905871348	0.847034407881696\\
1.59765539878615	0.868458948737302\\
1.71043164215823	0.887462728355275\\
1.82605235097095	0.904179145694992\\
1.94432016907157	0.918770756582008\\
2.06504723002923	0.931418084985431\\
2.18805606661672	0.942309896861698\\
2.31318012072368	0.951635245331063\\
};
\addlegendentry{\textsf{TCFIR}}

\end{axis}
\end{tikzpicture}%
    \caption{Typical $\gamma_0$ against the mutual information curves under OPRA and TCFIR transmission policies wherein the cut-off SNR $\gamma_0$ is constrained to the interval (0, 1]. Note that all-active relaying scheme is adopted with 4 relays. We assume the M\'alaga FSO fading with moderate turbulences ($\alpha=2.296,~\beta=1.822$).}
    \label{figure2}
\end{figure}
Clearly, $\gamma_0$ is a fixed point of the nonlinear equation (\ref{condition}). It has been demonstrated that $\gamma_0 \in (0,~1]$ irrespective of the number of relays or channel model employed and this follows from the properties of a CDF of a random variable, see Fig.~(\ref{figure2}). Therefore, (\ref{condition}) can be solved via iterative algorithms such as the Newton-Raphson method with a starting point conveniently selected from (0,~1].

\begin{figure}[H]
\centering
\setlength\fheight{5.5cm}
\setlength\fwidth{7.5cm}
%
%
\definecolor{mycolor1}{rgb}{0.00000,0.44700,0.74100}%
\definecolor{mycolor2}{rgb}{0.85000,0.32500,0.09800}%
\definecolor{mycolor3}{rgb}{0.92900,0.69400,0.12500}%
\definecolor{mycolor4}{rgb}{0.49400,0.18400,0.55600}%
\definecolor{mycolor5}{rgb}{0.46600,0.67400,0.18800}%
\begin{tikzpicture}

\begin{axis}[%
width=0.951\fwidth,
height=\fheight,
at={(0\fwidth,0\fheight)},
scale only axis,
xmin=0,
xmax=20,
xlabel style={font=\color{white!15!black}},
xlabel={\textsf{SNR (dB)}},
ymin=0,
ymax=3,
ylabel style={font=\color{white!15!black}},
ylabel={\textsf{Mutual Information (bits/s/Hz)}},
axis background/.style={fill=white},
axis x line*=bottom,
axis y line*=left,
legend style={at={(0.3,0.99)},legend cell align=left, align=left, draw=white!15!black,draw=none,fill=none}]

\draw [-latex,black,line width=.5pt] (15,.5) to (10,2);

\node[right, align=left, rotate=0]
at (axis cs:11,.35) {\scriptsize{\textsf{ $\delta = \left\{5, 1, 0.5, 0.1\right\}$ }}};

\addplot [color=blue, line width=1.3pt]
  table[row sep=crcr]{%
0	0.430173691135443\\
2	0.57914313651813\\
4	0.757837135225122\\
6	0.964971175354941\\
8	1.19792671467302\\
10	1.45325740421298\\
12	1.72720193405031\\
14	2.01609621411504\\
16	2.31664231902001\\
18	2.62604151787046\\
20	2.94202411684173\\
};
\addlegendentry{\textsf{ORA}}

\addplot [color=cornellred, line width=1.3pt]
  table[row sep=crcr]{%
0	0.425620138353231\\
2	0.572117301372774\\
4	0.74769910373012\\
6	0.951211726607284\\
8	1.1802422267556\\
10	1.43157823052835\\
12	1.70167456173248\\
14	1.98703369982267\\
16	2.28446196661879\\
18	2.59120552069227\\
20	2.90499238685507\\
};
\addlegendentry{\textsf{Effective}}

\addplot [color=cornellred, line width=1.3pt]
  table[row sep=crcr]{%
0	0.408070163607375\\
2	0.544981389331164\\
4	0.708316326314379\\
6	0.897255646633432\\
8	1.1100092374989\\
10	1.34414891883828\\
12	1.59693348403071\\
14	1.86557425476229\\
16	2.14742260457538\\
18	2.4400832491474\\
20	2.74146791241446\\
};
\addplot [color=cornellred, line width=1.3pt]
  table[row sep=crcr]{%
0	0.387607961097666\\
2	0.513311911107754\\
4	0.662054967288694\\
6	0.833099049978076\\
8	1.02501490198107\\
10	1.23594164474364\\
12	1.46381910393278\\
14	1.70656608384525\\
16	1.96220023399473\\
18	2.22890757383266\\
20	2.50507418904738\\
};
\addplot [color=cornellred, line width=1.3pt]
  table[row sep=crcr]{%
0	0.272632509159578\\
2	0.34054969883487\\
4	0.414844642083675\\
6	0.494224259314034\\
8	0.577475161118007\\
10	0.663557899473577\\
12	0.75164290921432\\
14	0.841104404264405\\
16	0.931491046213359\\
18	1.02248862371917\\
20	1.11388428786157\\
};
\end{axis}
\end{tikzpicture}%
    \caption{Channel capacity for different adaptive transmission schemes. Note that selective relaying mode has been adopted with 3 relays. We assume the Double Generalized Gamma FSO fading with weak turbulences ($\alpha_1=3,~\alpha_2=1.5$) under intensity modulation and direct detection. }
    \label{figure3}
\end{figure}
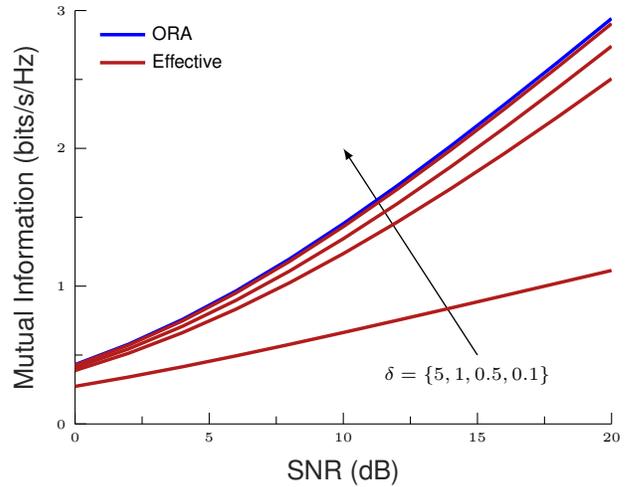
In the case of effective capacity, Fig.~(\ref{figure3}) suggests that as the product between $\phi,~B,~T_f$ (QoS exponent, bandwidth and fading block/frame length) decreases, the effective capacity increases to match the ergodic capacity.

\section{Conclusion}
In this paper, we have revisited the performance analysis of multihop regenerative relaying systems under the Generalized Fox $H$-function fading presented for different adaptive transmssion schemes. In particular, we presented the common fading models with the Fox $H$-function and in particular, we evaluated the mutual information for serial, all-active and selective relaying protocols and under for three FSO fading models; Gamma-Gamma, M\'alaga and Double Generalized Gamma, respectively. As a future direction, we plan to present the closed-form expressions of the coverage and mutual information for regenerative and non-regenerative multihop relaying under different power and rate adaptation schemes.

\bibliographystyle{IEEEtran}
\bibliography{main.bib}
\end{document}